\documentclass[10pt,letterpaper,twocolumn]{article} 

\usepackage{ol2}
\usepackage[draft]{hyperref}
\usepackage{amsmath}

\begin{document}

\twocolumn[

\title{Terahertz optically tunable dielectric metamaterials without microfabrication}

\author{Carlo Rizza $^{1,2}$, Alessandro Ciattoni$^{2,*}$, Lorenzo Columbo$^{1,3}$, Massimo Brambilla$^{3,4}$, Franco Prati$^{1,5}$}

\address{
$^1$Dipartimento di Scienza e Alta Tecnologia, Universit\`a  dell'Insubria, via Valleggio 11, I-22100 Como, Italy \\
$^2$Consiglio Nazionale delle Ricerche, CNR-SPIN, via Vetoio 1, I-67100 Coppito L'Aquila, Italy\\
$^3$Consiglio Nazionale delle Ricerche, CNR-IFN, via Amendola 173, I-70126 Bari, Italy\\
$^4$Dipartimento Interateneo di Fisica, Politecnico di Bari, via Amendola 173, I-70126 Bari, Italy\\
$^5$CNISM, Research Unit of Como, via Valleggio 11, I-22100 Como, Italy \\
$^*$Corresponding author: alessandro.ciattoni@aquila.infn.it}

\begin{abstract}
We theoretically investigate the terahertz dielectric response of a semiconductor slab hosting an infrared photoinduced grating. The periodic
structure is due to the charge carries photo-excited by the interference of two tilted infrared plane waves so that the grating depth and period
can be tuned by modifying the beam intensities and incidence angles, respectively. In the case where the grating period is much smaller than the
terahertz wavelength, we numerically evaluate the ordinary and extraordinary component of the effective permittivity tensor by resorting to
electromagnetic full-wave simulation coupled to the dynamics of charge carries excited by infrared radiation. We show that the photoinduced
metamaterial optical response can be tailored by varying the grating and it ranges from birefringent to hyperbolic to anisotropic negative
dielectric without resorting to microfabrication.
\end{abstract}

]

\noindent

Active metamaterials exhibiting an externally driven electromagnetic response are highly desirable for many interesting applications. In the
terahertz (THz) frequency range, many researchers have theoretically and experimentally considered a tunable response by semiconductor
inclusions in the metamaterial composite. In fact, the semiconductor permittivity can be modified by exploiting different excitation mechanisms
encompassing photocarrier injection \cite{Libon_1,Janke_1,Padilla_1,Shen_1,Rizza_1}, application of a bias voltage \cite{Chen_1} and/or thermal
carrier excitation \cite{Rivas_1,Sanchez_1,Chen_2}. Moreover, since a spatially modulated optical beam can induce in a semiconductor either
dielectric or metallic response to THz radiation \cite{Okada_1}, a complex effective metallo-dielectric structure can be achieved without
time-consuming microfabrication processes. For example, T. Okada et al. in Ref.\cite{Okada_1} considered a periodical grating created by an
optical pulse on a Si prism showing that the photo-induced structure displays metallic properties and, as a consequence, it is able to support
surface waves.

In this Letter, we suggest a novel class of infrared (IR) driven tunable dielectric metamaterials for THz radiation. Here, a periodic dielectric
grating is photo-generated by the interference pattern of two tilted IR plane waves with frequency within the semiconductor absorption band;
therefore, as opposed to standard photonic structures, the bulk grating is amenable to be reconfigured since one can mold the photo-induced
carrier density by changing the IR illumination. More precisely, the grating depth and period can be straightforwardly modified by changing the
intensity and incidence angle of the two driving IR plane waves, respectively. In the situation where the grating period is much smaller than
the THz wavelength the overall structure behaves as an homogeneous medium whose effective permittivity tensor can be evaluated through first
principle simulations. We show that the effective dielectric response is generally that of a uniaxial crystal with optical axis along the
grating and that, in a specific range of IR radiation intensity, its ordinary and extraordinary principal permittivities have different signs.
As consequence the considered structure can act as a hyperbolic material. This kind of metamaterials show very remarkable properties related to
the hyperbolic nature of their isofrequency contour such as, for example, subwavelength imaging through hyperlensing \cite{Andry_1}. Some of the
possible realizations of hyperbolic media in the THz frequency range include homogenous naturally-occurring materials (such as triglycine
sulfate \cite{Alek_1}), periodic array of aligned carbon nanotubes \cite{Nefedov_1} and multilayer graphene structure \cite{Iorsh_1}.

The proposed tunable photo-induced dielectric metamaterial is sketched in Fig. 1.
\begin{figure}
\center
\includegraphics*[width=0.5\textwidth]{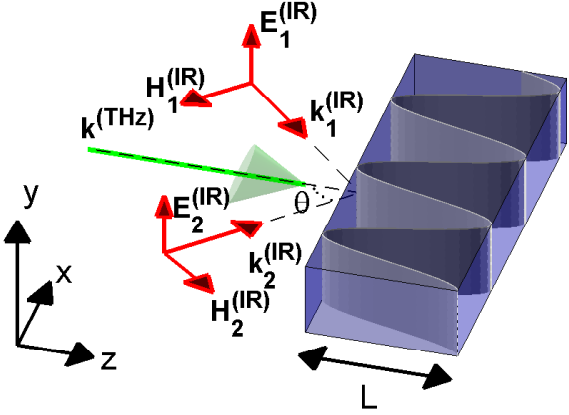}
\caption{(Color on-line) Sketch of tunable photo-induced THz dielectric metamaterial slab. The two infrared (IR) plane waves (with incidence
angles $\pm \theta$) induce a dielectric grating along the $x$-axis affecting the THz radiation. The terahertz (THz) plane wave normally
impinges on the semiconductor slab of thickness $L$.}
\end{figure}
A semiconductor infinite slab of thickness $L$ is illuminated by two IR tilted plane waves (at a frequency $\omega^{(\mathrm{IR})}$), namely
${\bf E}_j^{(\mathrm{IR})}(x,z,t)=E_0^{(\mathrm{IR})} \cos[ {\bf k}_j^{(\mathrm{IR})}\cdot {\bf r} -\omega^{(\mathrm{IR})}t ] \hat{{\bf e}}_y$
with $E_0^{(\mathrm{IR})}=\sqrt{I/(2 \epsilon_0 c)}$. Here ${\bf r}=x \hat{\bf e}_x + z \hat{\bf e}_z$, $I$ is the radiation intensity, ${\bf
k}_1^{(\mathrm{IR})} = k^{(\mathrm{IR})} (-\sin \theta \hat{\bf e}_x + \cos \theta \hat{\bf e}_z)$ and ${\bf
k}_2^{(\mathrm{IR})}=k^{(\mathrm{IR})} (\sin \theta \hat{\bf e}_x + \cos \theta \hat{\bf e}_z)$ where
$k^{(\mathrm{IR})}=\omega^{(\mathrm{IR})}/c$ and $\theta $ is the incidence angle and ($\epsilon_0$ and $c$ are the vacuum permittivity and
speed of light, respectively). The IR beams photo-excite electrons in conduction band and such excitation depends on the overall intensity
distribution and, as a consequence, plane waves interference spatially modulates the charge carrier density $N(x,z)$ along the $x$-axis. The
steady state profile of the photo-generated carrier density is given by \cite{Garmire_1}
\begin{equation}
\label{N-dyn} D \nabla^2 N -\frac{N}{\tau(N)} + \frac{\epsilon_0}{2 \hbar} Im[\epsilon^{(\mathrm{IR})}(N)]|E^{(\mathrm{IR})}|^2=0
\end{equation}
where $D$ is the diffusion coefficient, $\tau(N)$ is the density-dependent electron-hole recombination time, $\epsilon^{(\mathrm{IR})}(N)$ is
the density-dependent semiconductor permittivity at the IR frequency and $E^{(\mathrm{IR})}(x,z)$ is the amplitude of the monochromatic IR field
within the semiconductor slab ($\hbar$ is the reduced Planck constant). The recombination rate can be expressed by the polynomial approximation
$1/\tau(N)=A+BN+CN^2$ where the first term $A$ is the nonradiative recombination rate coefficient whereas the terms with $B$ and the $C$
describe radiative transitions and Auger recombinations, respectively \cite{Garmire_1}. The IR dielectric permittivity
$\epsilon^{(\mathrm{IR})}(N)$ can be described by adopting the microscopic model in Ref.\cite{Tissoni_1,Chow_1}. This model is obtained in the
quasi-equilibrium regime through the free-carrier approximation and considering some relevant effects such as the Urbach tail and the bandgap
renormalization. The incident THz field is a plane wave (at a frequency $\omega^{(\mathrm{THz})}$) normally impinging onto the semiconductor
slab interface (see Fig. 1), i.e. ${\bf E}^{(\mathrm{THz})}={\bf E}_0^{(\mathrm{THz})}\cos[k^{(\mathrm{THz})} z -\omega^{(\mathrm{THz})}t]$
where ${\bf E}_0^{(\mathrm{THz})}$ lies in the $x$-$y$ plane and $k^{(\mathrm{THz})}=\omega^{(\mathrm{THz})}/c$. We can described the
semiconductor THz dielectric response by the Drude dielectric model
\begin{equation}
\label{Drude} \epsilon^{(\mathrm{THz})}=\epsilon_{b}+\frac{i}{\epsilon_0} \frac{ e^2\tau/m^*N }{ \omega^{(\mathrm{THz})}
(1-i\omega^{(\mathrm{THz})}\tau) }
\end{equation}
where $\epsilon_{b}$ is the background dielectric constant, $e$ is the electron charge unit, $m^*$ is the effective mass of electrons in
semiconductor and $\tau$ is the free electron relaxation time. Note that by externally changing the intensity $I$ and the incidence angle
$\theta$ of IR illumination, the carrier density $N$ is locally modified (see Eq. (\ref{N-dyn})) and this in turn induces a spatial modulation
of $\epsilon^{(\mathrm{THz})}$ (see Eq. (\ref{Drude})).

The IR driven functionalities of the semiconductor slab were investigated through $2$D full-wave simulations performed with the comsol RF module
\cite{Comsol_1}.  THz linear and IR nonlinear Maxwell equations with $\epsilon^{(\mathrm{THz})}$ given by Eq. (\ref{Drude}) and
$\epsilon^{(\mathrm{IR})}$ calculated according to \cite{Tissoni_1,Chow_1} were coupled with Eq. (\ref{N-dyn}) for the carrier density profile.
The transmissivity and reflectivity coefficients of the slab were evaluated, from which its effective dielectric permittivity can be retrieved
\cite{Smith_1}.

In our simulations, we have set $L=2.5$ $\mu$m and $\theta=0.04$ rad, $\omega^{(\mathrm{IR})}=2168$ THz, $\omega^{(\mathrm{THz})}=18.85$ THz and
we have used typical parameters of GaAs bulk at room temperature i.e. $D=4 \cdot 10^{-3}$ m$^2/$s, $A=10^8 $ $1/$s \cite{Tissoni_1},  $B=7.2
\cdot 10^{-16} $ m$^3/$s, $C=10^{-42}$ m$^6/$s \cite{Varshni_1} and $\epsilon_{b}=13.32+i8.95 \cdot 10^{-3}$ \cite{Palik_1}, $\tau=3.23 \cdot
10^{-13}$ s, $m^*=0.067 m_0$ (where $m_0$ is the electron mass) \cite{Blakemore_1}. We have considered that two vacuum layers are placed at the
facets of the GaAs slab for providing IR and THz external illumination and for evaluating transmission and reflection coefficients. We have
adopted periodic boundary conditions along the $x$-axis for both THz and IR radiation and carrier density. Furthermore, at the entrance and the
exit facets (orthogonal to the $z$-axis) of the integration domain, we have used matched boundary conditions for THz and IR electromagnetic
fields, whereas at the entrance and exit facets of the GaAs slab, we have required $\partial_z N=0$. Note that the integration domain for $N$
coincides with the GaAs slab where photo-generation takes place.
\begin{figure}
\center
\includegraphics*[width=0.5\textwidth]{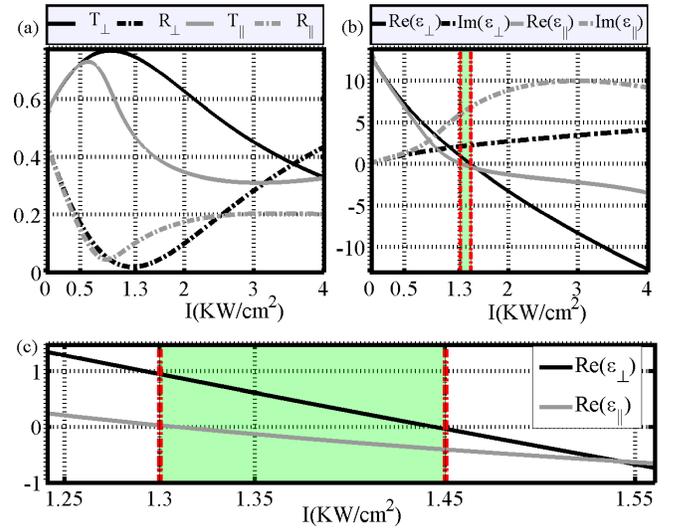}
\caption{(Color on-line) (a) Transmissivity $T_\perp$ ($T_{||}$) and Reflectivity $R_\perp$ ($R_{||}$) of  an ordinary (extraordinary) wave as a
function of the incident IR intensity $I$. (b) Effective dielectric permittivity ($\epsilon_{||}$,$\epsilon_{\perp}$) retrieved by
transmissivity and reflectivity of panel (a) as a function of the incident IR intensity $I$. (c) Magnification of the shadow region of panel (b)
where the THz response is hyperbolic.}
\end{figure}
\begin{table*}[htb]
  \centering
  \caption{Effective THz dielectric behavior of the considered GaAs slab at different IR intensity.}
  \begin{tabular}{ccccc}
  \\ \hline
                                   & Birefringent medium        & Hyperbolic medium            & Anisotropic negative dielectric \\
                                   & $0<I$ (KW/cm$^2$) $<1.3$   & $1.3<I$ (KW/cm$^2$) $<1.45$  & $I$ (KW/cm$^2$) $>1.45$ \\ \hline
$Re(\epsilon_{||})$                & $13.3-0$                   & $0--0.4$                     &  $<-0.4$   \\
$Im(\epsilon_{||})$                & $0-6.0$                    & $6.0-6.8$                    & $>6.8$     \\
$Re(\epsilon_{\perp})$             & $13.3-0.9$                 & $0.9-0$                      & $<0$ \\
$Im(\epsilon_{\perp})$             & $0-2$                      & $2-2.2$                      & $>2.2$     \\ \hline
  \end{tabular}
\end{table*}
Here, the period of the induced grating is $p=\pi c/(\omega^{(\mathrm{IR})} \sin \theta)=10.86$ $\mu$m $\simeq \lambda^{(\mathrm{THz})}/10$ so
that we assume the slab dielectric response to coincide with that of a homogeneous medium at THz frequencies. Within such a homogenized regime,
THz fields with linear polarizations parallel (${||}$-polarized) or orthogonal ($\perp$-polarized) to the $x$ axis generally have unequal
effective dielectric permittivities, $\epsilon_{||}$ and $\epsilon_{\perp}$, respectively. In Fig. 2(a) we report the slab transmissivity and
reflectivity of ${||}$- and ${\perp}$- polarized waves as functions of the IR intensity $I$ in the range $0$ KW/cm$^2$ $<I<$ $4$ KW/cm$^2$. In
Fig. 2(b) we plot the effective dielectric functions retrieved from the complex reflection and transmission coefficients \cite{Smith_1} whose
squared moduli are reported in panel (a). From Fig. 2(b) it is evident that we can steer the THz dielectric response by varying the intensity of
the two IR plane waves. Specifically, the considered structure exhibits three different behaviors, at different values of the IR intensity,
whose properties are summarized in Table 1. At low IR intensities ($I<$ $1.3$ KW/cm$^2$) the medium behaves as an uniaxial medium
($Re(\epsilon_{||}) \geq Re(\epsilon_{\perp})>0$). In a second IR intensity range $1.3$ KW/cm$^2$ $<I<$ $1.45$ KW/cm$^2$, the medium
birefringence increases to the point that $Re(\epsilon_{\perp})>0$ and $Re(\epsilon_{||})<0$, so that the effective permittivity tensor is
indefinite and the effective medium behaves as an hyperbolic medium. The IR intensity range where medium hyperbolicity occurs is highlighted
both in Fig. 2(b) and in Fig. 2(c) (which contains a magnified version of panel (b)) with a shadowed stripe. Finally, at higher IR intensities,
for $I>$ $1.45$ KW/cm$^2$, the structure effectively displays the behavior of an anisotropic negative dielectric metamaterial
($Re(\epsilon_{\perp}) <0$ and $Re(\epsilon_{||})<0$).

In conclusion, we have shown that a semiconductor slab has a THz dielectric effective behavior which can externally be tuned through suitable IR
illumination, it spanning from slight birefringence to anisotropic negative dielectric behavior and, in particulary, it is worth noting that the
semiconductor slab shows a hyperbolic response in a specific range of IR intensity. In addition to the multidisciplinary value of our results
ensuing from the mingling of semiconductor and metamaterial physics concepts, it is remarkable that our approach can easily be generalized to
more complex intensity patterns of the IR radiation yielding different and more exotic ways of achieving averaged effective medium responses.

This research has been funded by the Italian Ministry of Research (MIUR) through the "Futuro in Ricerca" FIRB-grant PHOCOS - RBFR08E7VA.

\end{document}